\documentclass[josa,twocolumn,showpacs,floatfix,superscriptaddress,10pt]{revtex4-1}

\usepackage{amssymb}
\usepackage[usenames,dvipsnames]{color}
\usepackage{graphicx}
\usepackage{subfigure}
\usepackage{epsfig}
\usepackage{amsmath}
\usepackage{subfigure}
\usepackage[normalem]{ulem}

\begin{document}

\title[Different operational meanings for CV entanglement criteria]
{Different operational meanings of continuous variable Gaussian
entanglement criteria and Bell inequalities}

\author{D. Buono}
\address{Dipartimento di Ingegneria Industriale, Universit\`{a}
degli Studi di Salerno, via Giovanni Paolo II, I-84084 Fisciano (SA), Italy}

\author{G. Nocerino}
\address{Trenitalia spa, DPR Campania, Ufficio di Ingegneria della
Manutenzione, IMC Campi Flegrei, Via Diocleziano 255, 80124 Napoli, Italy}

\author{S. Solimeno}
\address{Dipartimento di Fisica, Universit\`{a} "Federico II", Complesso
Universitario Monte Sant'Angelo, I-80126 Napoli, Italy}

\author{A. Porzio}
\address{CNR -- SPIN, Unit\`{a} di Napoli, Complesso Universitario Monte
Sant'Angelo, I-80126 Napoli, Italy}

\begin{abstract}
Entanglement, one of the most intriguing aspects of quantum mechanics,
marks itself into different features of quantum states. For this
reason different criteria can be used for verifying entanglement.
In this paper we review some of the entanglement criteria casted for
continuous variable states and link them to peculiar aspects of the original
debate on the famous EPR paradox. Moreover, we give a handy expression
for valuating Bell--type non--locality on Gaussian states. We also present
the experimental measurement of a particular realization of the Bell operator
over continuous variable entangled states produced by a sub--threshold
type--II OPO.
\end{abstract}

\maketitle

\section{Introduction}

Since the first reply by Schr\"{o}dinger \cite{Schrodinger1935} to the
famous EPR paper \cite{Einstein1935} the word "entanglement" has been
primarly used for indicating a class of quantum states that shows non--local
features. Discussing the dynamical properties of a composite system made of
two subsystems that, after mutual interaction, move away one from the other
Einstein Podolsy and Rosen concluded that quantum mechanics was not-complete
and that some more local (hidden) dynamical variables would have been
necessary for a correct description of the physical reality. Essentially,
they pointed to two quantum--mechanical aspects that they found
counter--intuitive. In primis, the possible ambiguity of the wave function
so that "\ldots\ as a consequence of two different measurements performed
upon the first system, the second system may be left in states with two
different wave functions \ldots\ ". The second aspect was later indicated,
by Einstein itself, as a \textit{spooky action at distance} "\ldots since at
the time of measurement the two systems no longer interact, no real change
can take place in the second system in consequence of anything that may be
done to the first system \ldots\ ". We nowadays know that they were wrong
and that quantum mechanics gives, so far, a complete representation of this
strange phenomenon.

Up to the late fifties of last century the debate on entanglement was mostly
confined to the fundamental aspect of quantum mechanics and the word itself
hadn't any particular operational meaning. In 1957 a paper by Bohm and
Aharonov \cite{Bohm1957} moved the focus from the original \textit{%
Gedankenexperiment} toward more feasible and intuitive physical
implementations and, in particular, to spin--like systems. This paved the
way to a more complete theoretical analysis of the hidden variables scenario
that leads to the famous Bell inequalities\footnote{%
In the original Bell's paper only one inequality is discussed, namely Eq.
(15) in Ref. \cite{Bell1964} that was, five years later, translated into a
set of experimentally verifiable inequalities, since then known as CSHS type
inequalities \cite{Clauser1969}.}\cite{Bell1964} for dichotomic quantum
variables. This made spin--like systems the preferential candidates for
proving the failure of any hidden variables hyphotesis. Single photons have
been then widely used in several experimental tests of the Bell inequalities
(see for a review Ref. \cite{Horodecki2009}). Very recently, a novel
experiment made the photon the first physical system for which each of the
main loopholes has been closed \cite{Giustina2013}.

On the other hand, the original formulation of the EPR paradox was based on
continuous variable systems. So that, in 1986 Reid and Walls proposed the
first translation of Bell inequalities into the continuous varaible language 
\cite{Reid1986} and in 1992 the first experimental realization of an
EPR--like system appeared \cite{Ou1992PRL}. Since then, a few more attempts
have been carried out for translating the Bell argument into the language of
continuous variables (CV) \cite{Braunstein2005}. Among them the one proposed
by Banaszek and W\'{o}dkiewicz \cite{Banaszek1998} considers the relation
between the Wigner function of the state and non--locality.

Entanglement, in its orignal formulation, states the existence of global
states of a composite system which cannot be written as a product of states
of individual subsystems. While this definition set an univoque border
between separable and entangled states, entanglement gives rise to different
features of quantum systems and can be seen under different perspectives 
\cite{Wiseman2007}. On one hand, mathematically defining entanglement as a
property of the composite system wavefunction, make it intrinsically related
to pure states \cite{Cavalcanti2005}. On the other hand, we all know that
experimentally acessible states are mixed, so that feasible entanglement
tests have to be related to density matrices rather than wave functions \cite%
{Werner1989}.

In this paper we aim at discussing the operational meanings of different
criteria usually employed for assessing CV entanglement. In particular, we
will link each of them to different facets of the original entanglement
debate. We will apply them to entangled Gaussian states (GS) \cite%
{Weedbrook2012,Olivares2012} produced by a type--II sub--threshold frequency
degenerate OPO. By experimentally analysing the properties of experimentally
generated CV entangled states we will express all these criteria in terms of
the covariance matrix elements. Moreover, we give a novel handy relation
that describes in a simple way the connection among entanglement, purity and
Bell's non-locality. The experimental analysis prove that some entanglement
features are strongly hold against decoherence while Bell inequality is
violated only for nearly pure states.

The paper is organized as follows. In Sect. II the properties of Guassian
states are reviewed. Then, in Sect. III, a summary of different entanglement
criteria is given. Each of the presented criteria is related to a particular
feature of entanglement. In Sect. IV a Bell-type inequality is given in
terms of the Gaussian states properties. While in Sect. V a discussion on
the effects of decoherence on states violating Bell inquality is presented.
In Sect. VI we report an overview on some experimental results and, in
particular, an \textit{a posteriori} Bell test on an effective bipartite CV
entangled state. Eventually, in Sect. VII conclusions are drawn.

\section{Gaussian states}

A continuous-variable bi--partite GS is a two-mode state, on the Hilbert
space $\mathcal{H}=\mathcal{H}_{a}\otimes \mathcal{H}_{b}$, whose
characteristic function or, equivalently, Wigner function in phase space is
Gaussian:%
\begin{equation}
W(\mathbf{K})=\frac{\exp \{-\frac{1}{2}\mathbf{K}^{T}\mathbf{\sigma }^{-1}%
\mathbf{K}\}}{2\pi \sqrt{Det[\mathbf{\sigma }]}}  \label{Wigner}
\end{equation}%
where $\mathbf{K\equiv }\left( X_{a,\vartheta },~X_{a,\vartheta +\pi
/2},X~_{b,\vartheta },~X_{b,\vartheta +\pi /2}\right) $ is the vector of a
set of orthogonal quadratures, for mode $a$ and $b$ respectively. (being $%
\widehat{X}_{k,\vartheta }\equiv \frac{\widehat{a}e^{i\vartheta }+\widehat{a}%
^{\dag }e^{-i\vartheta }}{\sqrt{2}}$). We remind that the pair $\widehat{X}%
_{0}=\widehat{X},$ and $\widehat{X}_{\pi /2}=\widehat{Y}$ ($\left[
X_{k},Y_{k}\right] =i$) associated to a single \textit{e.m. }mode is the
analogue to the position/momentum pair for a mechanical oscillator. This
makes optical mode a good candidate for replicating EPR states in their
original fashoin. All the features of GS are embedded in the second order
momenta of the joint quadrature distribution, namely the covariance matrix%
\footnote{%
It is possible to define a CM for any state but only in the case of GS it is
exhaustive for describing the state.} (CM) $\mathbf{\sigma }$ in Eq. (\ref%
{Wigner}). A pure GS can be seen as the action of a displacement and a
squeezing operator onto the vacuum state. While, the most general mixed GS
can be obtained replacing the vacuum with a thermal field at finite
temperature.

For a bipartite state the $\mathbf{\sigma }$ is a $4\times 4$ matrix, with
elements $\sigma _{hk}\equiv \frac{1}{2}\left\langle \left\{
K_{k},K_{h}\right\} \right\rangle -\left\langle K_{k}\right\rangle
\left\langle K_{h}\right\rangle $ (being $\left\{ K_{k},K_{h}\right\} \equiv
K_{k}K_{h}+K_{h}K_{k}$ the anti--commutator). $\mathbf{\sigma }$ can be
written in the form%
\begin{equation}
\mathbf{\sigma =}\left( 
\begin{array}{cc}
\boldsymbol{\alpha } & \boldsymbol{\gamma } \\ 
\boldsymbol{\gamma }^{\top } & \boldsymbol{\beta }%
\end{array}%
\right) ,  \label{CM}
\end{equation}%
where $\boldsymbol{\alpha }$ and $\boldsymbol{\beta }$ represent
self--correlation of the single subsystem and $\boldsymbol{\gamma }$
describes the cross correlation between the two subsystems. Remarkably, a GS
represents any quantum system whose evolution can be described by a at least
bilinear Bosonic Hamiltonian. In particular this is the case of optical
parametric oscillators (OPO).

Any CM, representing a physical state, can be transformed into the so-called
standard form \cite{Duan2000} 
\begin{equation}
\mathbf{\sigma }_{S}=\left( 
\begin{array}{cccc}
n & 0 & c_{1} & 0 \\ 
0 & n & 0 & c_{2} \\ 
c_{1} & 0 & m & 0 \\ 
0 & c_{2} & 0 & m%
\end{array}%
\right) ~.  \label{CM initial}
\end{equation}%
by means of local symplectic transformations\footnote{%
A transformation is symplectic if it preserves the metric. From a physical
point of view it means that it preserves canonical commutation relations.}
where $n$, $m$, $c_{1}$ and $c_{2}$ are determined by four local symplectic
invariants $I_{1}\equiv \det (\mathbf{\alpha })=n^{2}$, $I_{2}\equiv \det (%
\mathbf{\beta })=m^{2}$, $I_{3}\equiv \det (\mathbf{\gamma })=c_{1}c_{2}$, $%
I_{4}\equiv \det ({\boldsymbol{\sigma }})=\left( nm-c_{1}^{2}\right) \left(
nm-c_{2}^{2}\right) $. As a matter of fact, a sub--threshold type--II OPO,
due to the symmetry of its Hamiltonian, can only produce states whose CM is
a standard form \cite{PRA2012}.
Hereafter, whenever we refer to CMs we
will mean such standard form.
Moreover, at the time of birth inside the
non--linear crystal, the bipartite state we would analyse in the following
shows $n=m$ and $c_{1}=-c_{2}=c$.

From the symplectic invariants it is possible to give a criteria for
disitnguishing among physical and non-physical CMs. ${\boldsymbol{\sigma }}$
describes a physical state iff%
\begin{equation}
I_{1}+I_{2}+2I_{3}\leq 4I_{4}+\frac{1}{4}~.
\label{Heis rel. sympl invariant}
\end{equation}
We also note that a pure GS is a minimum uncertainty state and that the CM
relative to a pure state necessary has $\det ({\boldsymbol{\sigma }}%
)=I_{4}=1/16$ so that for a pure state%
\begin{equation}
c=\sqrt{n^{2}-1/4}~.  \label{c_pure}
\end{equation}%
while, for mixed symmetric states, $c<\sqrt{n^{2}-1/4}$. In general, for a
bipatite GS, the purity reads%
\begin{equation}
\mu \left( \sigma \right) =\frac{1}{4\sqrt{Det\left[ \sigma \right] }}.
\label{GaussPur}
\end{equation}

\section{Entanglement criteria}

A quantitative measure of entanglement for a mixed state is, so far, an
unsolved issue. This is probably due to the different operational
implications that different levels of quantum correlation open. At the same
time, there exist different necessary and/or sufficient conditions to asses
whether a given state is entangled or not. These criteria are easily
translated into experimental tests for entanglement. Here we aim to look at
the different criteria and connect them, logically, to the debate on the
original EPR paper \cite{Einstein1935}.

\subsection{\textit{Un--separability} criteria: PHS and Duan}

The first criterion was developed by considering the definition of entangled
states: a state of a composite system whose wavefunction cannot be given as
product of sub-systems wavefunction. Or, in the case of mixed states,
following the Werner extension to the relative density matrix \cite%
{Werner1989}. In the bi--partite case a density matrix represent a separable
state iff its can be written as a convex combination of the tensor product
of density operators relative to the two sub--systems%
\begin{equation}
\rho =\sum_{j}p_{j}\rho _{j1}\otimes \rho _{j2},  \label{SepSt}
\end{equation}%
where $\sum_{j}p_{j}=1$ while $\rho _{ji}$ $i=1,2$ are the density matrices
of subsystems $1$ and $2$. The criterion can be casted considering that if
one performs a partial transposition (\textit{i.e.} transposition of the
density matrix with respect to only one of the two Hilbert subspaces) $\rho $
transform into $\rho _{PT}$ that, for a state written in form given in Eq. (%
\ref{SepSt}) will still represent a physical state of the composite system.
Conversely, if the state un--separable, the tranformed density operator $%
\rho _{PT}$ would have no more a physical counterpart. This criterion, is
sometime referred to as the \textit{ppt} criterion (\textit{positivity}
under \textit{partial transposition}) or PHS from the names of the people
that proposed it for discrete (Peres \cite{Peres1996} and Horodecki \cite%
{Horodecki1997}) and continuous variables (Simon \cite{Simon2000}).
Translated into the CM language a bi--partite Gaussian state is separable
iff 
\begin{equation}
n^{2}+m^{2}+2\left\vert c_{1}c_{2}\right\vert -4\left( nm-c_{1}^{2}\right)
\left( nm-c_{2}^{2}\right) \leq \frac{1}{4}\text{ },  \label{PHScr}
\end{equation}%
and it is entangled otherwise. We also note that the PHS criterion is
invariant under symplectic transformations and that for pure states the
inequality is saturated.

The PHS criterion set, then, is roots into the fact that two systems that
have interacted cannot, even if the split apart after the interaction, be
described independently.

A second crierion, the Duan one \cite{Duan2000}, has been derived
considering that in presence of an entangled state also the Heisenberg
uncertainty principle, written for the joint system and a pair of EPR-like
operators, has to take into account the inherent quantum correlation. The
Duan criterion, a necessary condition for entanglement, for a CM in the
usual standard form reads:%
\begin{equation}
\sqrt{\left( 2n-1\right) \left( 2m-1\right) }-\left( c_{1}-c_{2}\right) <0~.
\label{DUANcr}
\end{equation}
Based on the calculation of the total variance of a pair of Einstein-
Podolsky-Rosen (EPR) type operators. It relies on the fact that the inherent
correlation reduces the total variance that, in separable states, is greater
than the sum of the standard quantum limit applied to the single subsystem.
In the case of
frequency degenerate type--II OPO, this result in the squeezing of the modes
obtained by letting the two entagled companions interfere \cite{PRL2009}.

\subsection{The EPR "Reid" criterion}

A stronger bound can be found by considering the original EPR \textit{%
Gedankenexperiment} where the paradox was found in the possibility of
determining the state of a far-away system by measuring its entangled
companion. For this reason this criterion is usually indicated as the EPR
criterion and was firstly introduced by Reid in 1989 \cite{Reid1989PRA1}, in
the very early days of quantum information. It describes the ability to
infer the expectation value of an observable on a sub--system by performing
a suitable measurement on the second sub--system. This criterion sets only a
sufficient condition for assessing entanglement being a stricter condition
on the strenght of quantum correlation. It can be easily given in terms of
CM elements: 
\begin{equation}
n^{2}\left( 1-\frac{c_{1}^{2}}{nm}\right) \left( 1-\frac{c_{2}^{2}}{nm}%
\right) <\frac{1}{4}~.  \label{EPRcr}
\end{equation}%
While the criterion is asymmetric under the exchange of the two sub-systems
so that the two definitions can make it ambiguous if one of the relations is
not satisfied. This is not the case of balanced systems ($m=n$) where this
one-side violation is not possible. The asymmetry of the criterion allows to
use it for the so called steering capability: the state of a far away system
can be steered by a suitable measurement on its entangled companion \cite%
{Handchen2012}.

\section{Bell--like inequality (non--locality) in phase space}

Generally, if we want to evaluate the non-locality of a state through a CHSH
inequality \cite{Clauser1969}, the operative form of the Bell one, we should
build a Bell operator representing a combination of dichotomic (true-false)
measurements. Then, if the expectation value of such a Bell operator
violates the corresponding inequality, the system is not considered local,
otherwise it would admit a classical description in terms of hidden
variables. A parity operator is dichotomic. It can be constructed, on the
photon number, for assigning $+1$ or $-1$ depending on whether an even or an
odd number of photons has been registered. In Refs. \cite%
{Banaszek1998,Banaszek1999PRL} a connection between the Wigner function of
the state and the joint measurement of the parity operator performed on the
bi--partite quantum state has been shown.

Here we want to give a handy expression that relates such a measurement to
the CM of a generic GS. We consider the Bell operator in the form given in
Eq. (7) of Ref. \cite{Banaszek1998}. The Bell type function $\mathcal{B}$\
is then, given by the linear combination of four expectation values%
\begin{equation}
\mathcal{B}\mathcal{=}\left\langle \mathfrak{W}(\mathbf{0},\mathbf{0}%
)\right\rangle +\left\langle \mathfrak{W}(\sqrt{\mathcal{I}},\mathbf{0}%
)\right\rangle +\left\langle \mathfrak{W}(\mathbf{0},-\sqrt{\mathcal{I}}%
)\right\rangle -\left\langle \mathfrak{W}(\sqrt{\mathcal{I}},-\sqrt{\mathcal{%
I}})\right\rangle ~,  \label{Bell}
\end{equation}%
where 
\begin{equation}
\left\langle \mathfrak{W}\left( \alpha _{1},\alpha _{2}\right) \right\rangle
\equiv \frac{\pi ^{2}}{4}W\left( \alpha _{1},\alpha _{2}\right) ~;
\label{corr}
\end{equation}%
with $W\left( \alpha _{1},\alpha _{2}\right) $ the Wigner function of the
state calculated in $\left( \alpha _{1},\alpha _{2}\right) $ and where $%
\alpha _{k}$ are complex amplitudes (and so is $\sqrt{\mathcal{I}}$ is Eq. (%
\ref{Bell})). Local theories, admitting a description in terms of local
hiddden variables, set the bound%
\begin{equation}
\left\vert \mathcal{B}\right\vert \mathcal{\leq }2.  \label{IneqFS}
\end{equation}

On one hand, any Bell inequality concerns the analysis of joint
probabilities measured at space--time--separated locations. So that, to
actually perform a Bell measure we should need to make repeated simultaneous
measurements at different space--time--separated locations stochastically
changing the detector settings (in this case the amounts of displacement).
Then by statistical analysis we could conclude or not the violation of the
CHSH type inequality.

On the other hand, Eq. (\ref{corr}) show that the knowledge of the Wigner
function, i.e. the full reconstruction of the quantum state gives an insight
to the local/non--local character of the state. Without running into
delicate questions we wish to show that, being a GS fully described by a
rather simple object, the CM, it is possible to evaluate, \textit{a
posteriori}, $\mathcal{B}$ on the state so to asses whether or not it is
Bell correlated without the need of reconstructing the whole Wigner
function. This paves the way to a handy experimental procedure to
disciminate among different levels of quantum correlations.

\subsection{Bipartite Gaussian state case}

Now, we consider the bipartite GS generated by a type--II OPO described by
the covariance matrix (\ref{CM initial}) $\mathbf{\sigma }$, with $n=m$ and $%
c_{1}=-c_{2}=c$.

It can be esaily found that the quantity (\ref{Bell}) becomes%
\begin{equation}
\mathcal{B}\left( \mathcal{I},n,c\right) =\frac{1+2\exp \left\{ -\frac{n}{%
n^{2}-c^{2}}\mathcal{I}\right\} -\exp \left\{ -\frac{n+c}{n^{2}-c^{2}}2%
\mathcal{I}\right\} }{4\left( n^{2}-c^{2}\right) }.  \label{BellGS}
\end{equation}%
The Bell function $\mathcal{B}\left( \mathcal{I},n,c\right) $ depends on the
state properties ($n,c$) and on a free parameter ($\mathcal{I}$). To look
for the maximum violation for a given state we need to look for the value of
the displacement amplitude $\mathcal{I}$ that nullifies the derivative $%
\frac{\partial \mathcal{B}\left( \mathcal{I},n,c\right) }{\partial \mathcal{I%
}}=0$. The maximum is, then, obtained for%
\begin{equation}
\widetilde{\mathcal{I}}\left( n,c\right) =\frac{n^{2}-c^{2}}{n+2c}\ln \left[ 
\frac{n+c}{n}\right] 
\end{equation}%
So that $\mathcal{\widetilde{B}}=\mathcal{B}\left|_%
{\mathcal{I}=\widetilde{\mathcal{I}}}\right.$ reads%
\begin{equation}
\mathcal{\widetilde{B}}\left( n,c\right) =\frac{1}{4\left( n^{2}-c^{2}\right) }%
\left[ 1+2\left( \frac{n+c}{n}\right) ^{-\frac{n}{n+c}}-\left( \frac{n+c}{n}%
\right) ^{-2\frac{n+c}{n+2c}}\right]   \label{B}
\end{equation}%
This gives the expectation value of the maximum value of the Bell operator $%
\mathcal{\widetilde{B}}$ as a function of the Gaussian state parameters. So
that, being possible to experimentally retrieve the CM of such a state \cite%
{PRL2009}, this formula can be used to perform an \textit{a posteriori} test
on the non--local property of the state.

Moreover, it is possible to relate $\mathcal{\widetilde{B}}$ to the purity of
the single subsystem $\mu _{s}\equiv \mu _{a}=\mu _{b}=1/\left( 2n\right) $.
Having in mind the interconnection between entanglement and the purity of the
constituent sub--systems \cite{Adesso2004}, we have:%
\begin{equation}
\mathcal{\widetilde{B}}\left( \mu _{s},C_{ab}\right) =\frac{\mu _{s}^{2}}{%
1-C_{ab}^{2}}\left[ 1+\left( 1+2C_{ab}\right) \left( 1+C_{ab}\right) ^{-2%
\frac{1+C_{ab}}{1+2C_{ab}}}\right]   \label{BwC}
\end{equation}%
where 
\begin{equation}
C_{ab}\equiv \frac{\left\langle \Delta X_{a}X_{b}\right\rangle }{\sqrt{%
\left\langle \Delta X_{a}^{2}\right\rangle \left\langle \Delta
X_{b}^{2}\right\rangle }}=\frac{c}{n}
\end{equation}%
is the correlation coefficient whose limit for a pure state is $%
C_{ab}^{2}=1-1/(4n^{2})=1-\mu _{s}^{2}$

It can also be proved that, when the Gaussian state is pure, $\mathcal{B}_{%
\widetilde{\mathcal{I}}}$ can be considered an entanglement witness: any
entangled state violates the Bell inequality and viceversa.

\subsection{Purity, entanglement and non-locality}

\begin{figure}[tph]
\centering\includegraphics[width=0.45\textwidth]{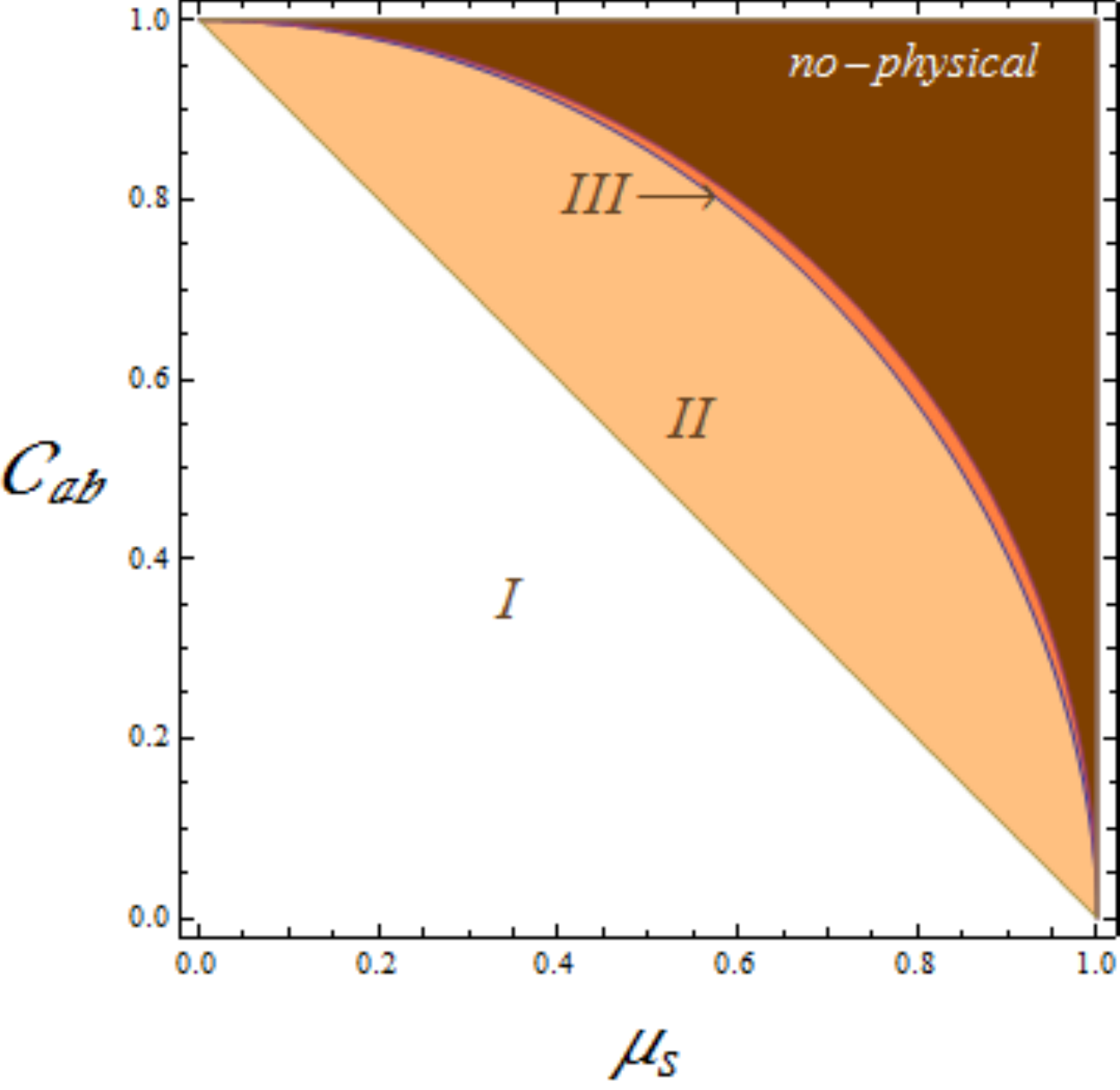}
\caption{Region plot of $\mathcal{\widetilde{B}}$ as function of purity $\protect%
\mu _{s}$ and the correlation coefficient $C_{ab}$. For the different
meanings of the plor regions see text.}
\label{BvsCmuS}
\end{figure}

For mixed states the above equivalence does not hold. Given a mixed system,
one has $\mu _{s}^{2}<1-C_{ab}^{2}$. So that, for a given correlation
coefficient $C_{ab}$, it is possible to set three boundaries for the values
of $\mu _{s}$%
\begin{eqnarray}
\mu _{D} &=&1-C_{ab}\text{,}  \notag \\
\mu _{B} &=&\left[ \frac{2\left( 1-C_{ab}^{2}\right) }{1+\left(
1+2C_{ab}\right) \left( 1+C_{ab}\right) ^{-2\frac{1+C_{ab}}{1+2C_{ab}}}}%
\right] ^{1/2}\text{,}  \notag \\
\mu _{P} &=&\left[ 1-C_{ab}^{2}\right] ^{1/2}.
\end{eqnarray}%
so that $\mu _{s}<\mu _{D}$ denotes separable states, $\mu _{D}<\mu _{s}<\mu
_{B}$ indicates entangled states that do not violate the Bell inequality,
and, finally, for $\mu _{B}<\mu _{s}<\mu _{P}$ the states are entangled and
violate the Bell inequality. For $\mu _{s}>\mu _{P}$ the relative CM would
be not physical. In this way it is possible to distinguish three (physical)
regions in the plane $\left( \mu _{s},C_{ab}\right) $ (see Fig. \ref{BvsCmuS}%
)

\begin{description}
\item[\textbf{Region} $I)$] Separable states compatible with the local
hidden variables theory ($\mathcal{\widetilde{B}}\left( \mu _{s},C_{ab}\right) <2
$).

\item[\textbf{Region} $II)$] Entangled states compatible with the local
hidden variables theory ($\mathcal{\widetilde{B}}\left( \mu _{s},C_{ab}\right) <2
$).

\item[\textbf{Region} $III)$] Entangled states not compatible with the local
hidden variables theory ($\mathcal{\widetilde{B}}\left( \mu _{s},C_{ab}\right) >2
$).
\end{description}

It is clear that there aren't separable GS that violate the Bell's
inequality. Instead, a state compatible with a local theory (\emph{i.e.}
compatible with a theory in hidden variables) can also be entangled. This
confirms the existence of different forms of quantum correlations and
non--locality.

\section{Gaussian noise does not break the entanglement, but it breaks the
Bell's nonlocality}

In this Section we want to analyze the behaviour of the Bell's nonlocality
subjected to passive Gaussian noise. We will see that when a pure ($c=\sqrt{%
n^{2}-1/4}$), entangled and Bell's non--local (i.e.$\mathcal{B}_{\widetilde{%
\mathcal{I}}}>2$) state evolves through a Gaussian channel, retains its
entanglement, but looses its Bell's nonlocality. This means that although,
at the time of its birth, the state is pure, so that it violates Bell's
inequality and breaks the Duan bound, decoherence highlights the different
nature of the two markers: the Gaussian state becomes local (in according to
Bell), i.e. it would admit a description in terms of local hidden variables,
although it remains entangled.

A bipartite state, described by the $CM$ Eq. (\ref{CM initial}), subjected
to the action of a passive Gaussian channel, undergoes a transformation such
that: \cite{PRA2012}%
\begin{eqnarray}
n_{T} &=&\frac{1-T}{2}+Tn_{1},  \notag \\
c_{T} &=&Tc_{1},
\end{eqnarray}%
where $n_{1}$ and $c_{1}$ are the $CM$ elements of the initial pure state
and $T$ is the trasmittivity of a fictituous beam splitter mimecking a lossy
transmission \cite{JPB2006}.

We can calculate the evolution of the Bell function $\mathcal{\widetilde{B}}%
\left( n_{1},c_{1}\right) $ (\ref{B}) starting from an initially pure state,
described by the $CM$ elements $n_{1}$ and $c_{1}$ and analyzing $\mathcal{%
\widetilde{B}}\left( n_{T},c_{T}\right) $ as a function of the coefficient of
transmissivity $T$ ($0<T<1$). So the Bell's function $\mathcal{\widetilde{B}}$
becomes a function $\mathcal{\widetilde{B}}_{T}$\ depending on the initial
(pure) state and transmissivity $T$\ of the channel. The relative
expression, ideed rather long and complicate, will be used for evaluating
the correspondence among experimental results and theory in the next section.

\section{Experimental results}

In a recent paper (see Ref. \cite{PRA2012}) we have analysed how
different entanglement and quantum signatures evolves under decoherence. In
this paper we wish to include the experimental analysis of the non--local
character under decoherence. We have to stress that, in view of the restrict
region ($T>90\%$) where one could expect a Bell inequality violation, we
have not observed any Bell inequality violation. This is essentially due to
the maximum overall transmission we can get from the OPO cavity to the
homodyne detector (63\%). Moreover, we stress that this is rather an a
posteriori check of the non-local character of the state than a Bell measure.

The analysed state is the one outing a sub--threshold type--II OPO \cite%
{APB2008}. The block scheme of the experiment is given in Fig. \ref%
{Fig:setup}. The full covariance matrix is retrieved by a single homodyne
detector \cite{JOB2005} following the procedure described in details in Ref. 
\cite{JOSAB2010}.

\begin{figure}[tph]
\centering\includegraphics[width=0.45\textwidth]{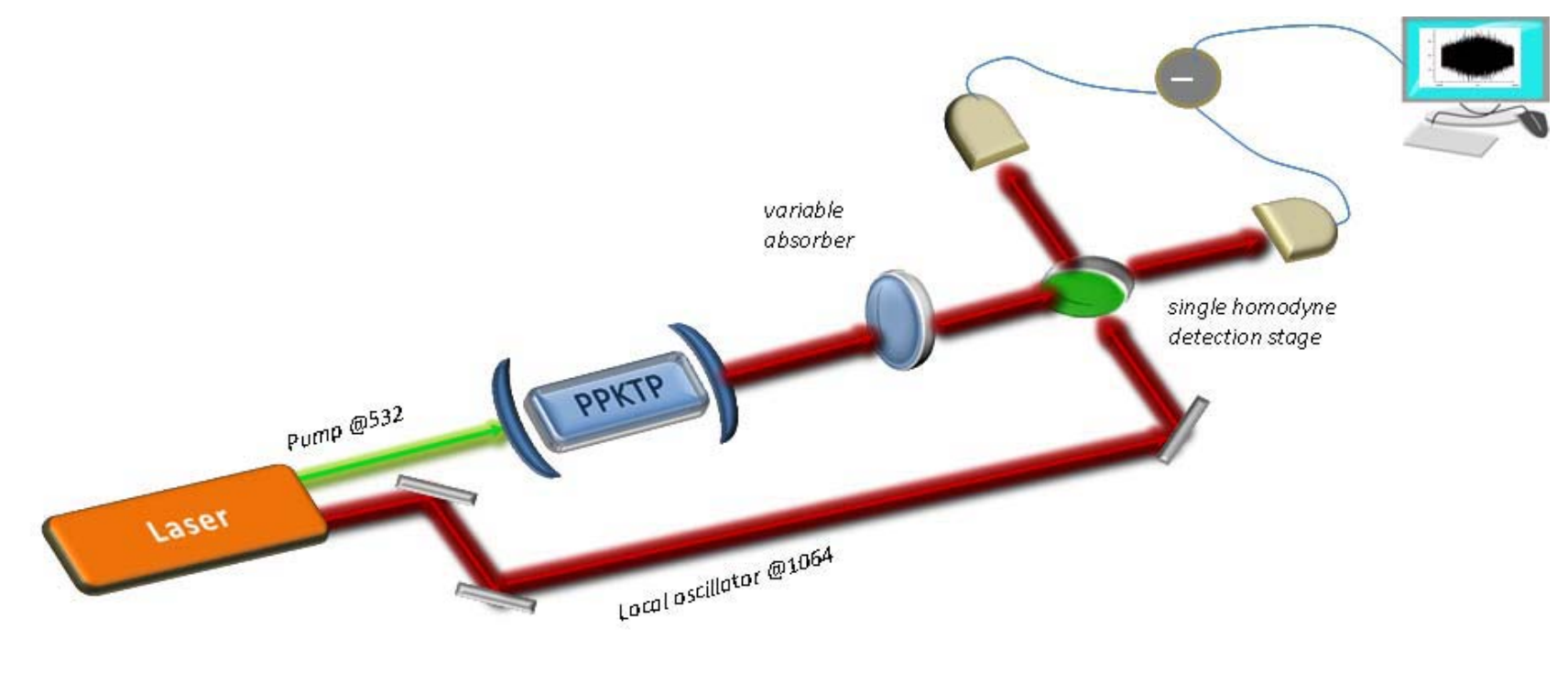}
\caption{Schematic representation of the experimental setup.
The generation stage is a
type--II OPO operating below the oscillation threshold. At the OPO output a
neutral adsorber mimicks the transmission over a real channel. The state is
reconstructed exploiting data collected by a single homodyne detector.}
\label{Fig:setup}
\end{figure}

\begin{figure}[!tph]
\centering\includegraphics[width=0.45\textwidth]{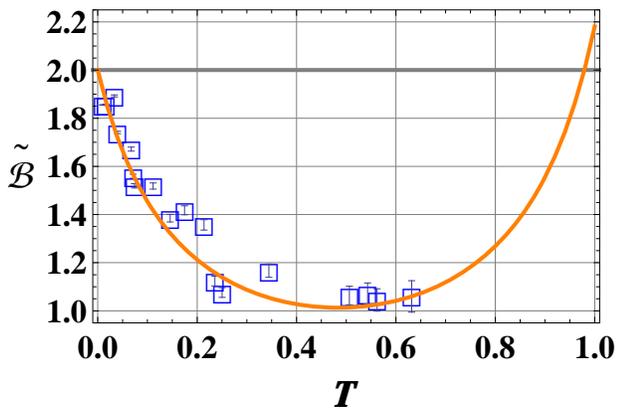}
\caption{Experimental evolution of the Bell's function vs. $T$, the transmittivity
of a variable absorber mimicking a realistic
transmission channel.}
\label{Fig:BellvsT}
\end{figure}

In Fig. \ref{Fig:BellvsT} we have plotted the experimental value obtained
for $\mathcal{\widetilde{B}}_{T}$. The continuous line represent the theoretical
expections for the pure ancestor state.

As it can be seen experimental data are in good agreement with the expected
evolution.

\section{Conclusions}

Different bounds have been, so far, discussed in
the literature for discriminating continuous variable separable and
entangled states. Each of them looks at slightly different facets of the
EPR paradox. So that, in this paper, they are presented in connection
to the original EPR arguments.

Moreover, for the first time, we express by a handy and direct formula 
a Bell--type inequaly, written for CV states, in terms of the covariance
matrix of a Gaussian state. We discuss its relation with the purity of the
entangled sub-systems and analyse, also experimentaly, its behaviour under
decoherence. So doing we have proved, experimentally, that, also in CV regime,
there exists mixed entangled states that do not violate the Bell inequality.
While, for pure states, any entangled state is Bell non--local.

\end{document}